\documentclass[10pt,twocolumn]{article}

\usepackage{amsmath}
\usepackage{isotope}
\usepackage{lipsum} 		% for filler text
\usepackage{blindtext} 		% Package to generate dummy text throughout this template 
\usepackage{siunitx}
\usepackage{multirow}
\usepackage{makecell}
\usepackage[perpage]{footmisc}

\usepackage[superscript, biblabel]{cite}  	% make citations be superscripts

\makeatletter
\renewcommand\@biblabel[1]{#1.}
\makeatother

\usepackage{etoolbox}
\patchcmd{\thebibliography}{\section*{\refname}}{}{}{}

\usepackage[margin=1.0in,hmarginratio=1:1,top=32mm,columnsep=15pt]{geometry}

\usepackage{color}				% Change the font color
\definecolor{dark-gray}{gray}{0.1}	% a dark gray
\usepackage{times}				% times text for the document times, san-serf is arial like
\usepackage{microtype} 			% Slightly tweak font spacing for aesthetics
\usepackage[english]{babel} 		% Language hyphenation and typographical rules

\usepackage[hang, font={footnotesize}, labelfont={bf,up}, textfont={sf,up}]{caption} 
\usepackage{booktabs} 	% Horizontal rules in tables
\usepackage{mwe}  		% blank figure
\usepackage{graphicx}  		% blank figure

\usepackage{enumitem} 		% Customized lists
\setlist[itemize]{noitemsep} 	% Make itemize lists more compact

\usepackage{abstract} % Allows abstract customization
\AtBeginDocument{}  		% removes abstract name

\usepackage{titlesec} 			% Allows customization of titles
\renewcommand\thesection{\Roman{section}.} 		  		%  section numbering 
\renewcommand\thesubsection{\thesection\Alph{subsection}.} 	% no subsection number
\renewcommand\thesubsubsection{\thesubsection\arabic{subsubsection}.} %subsubsection
\titleformat{\section}[block]{\normalfont\sffamily\bfseries}{\thesection}{1em}{\MakeUppercase}{} 	% Change the look of the section titles
\titleformat{\subsection}[block]{\normalfont\sffamily\bfseries}{\thesubsection}{1em}{}{}  % Change the look of the section titles
\titleformat{\subsubsection}[block]{\normalfont\sffamily\bfseries}{\thesubsubsection}{1em}{}{}  % Change the look of the section titles
\titlespacing*{\section}{0.0em}{1em}{0.25em}		% changing the spacing
\titlespacing*{\subsection}{0.0em}{1em}{0.25em}	% changing the spacing
\usepackage{indentfirst}						% indent first paragraph in the section

\usepackage{tocloft}% http://ctan.org/pkg/tocloft
\usepackage{fancyhdr} % Headers and footers

\fancypagestyle{plain}{
\fancyhf{}
\lhead{\color{dark-gray}\textit{}} 
\rhead{} 
}

\usepackage{titling} 		% Customizing the title section

\usepackage{hyperref} 	% For hyperlinks in the PDF
\hypersetup{
	backref=true,       
    	pagebackref=true,               
    	hyperindex=true,                
    	colorlinks=true,                
    	breaklinks=true,                
    	urlcolor= black,                
    	linkcolor=black,                
    	bookmarks=true,                 
    	bookmarksopen=false,
    	filecolor=black,
    	citecolor=blue
}

\pretitle{\begin{center}\large\bfseries} 	% Article title formatting
\posttitle{\end{center}} 				% Article title closing formatting
\title{\vspace{-0.3in} \sffamily{ 
 DT fusion through the $^5$He 3/2$^+$ ``Bretscher state'' accounts  for $\geq$25\% of our existence via nucleosynthesis
and for the possibility of fusion energy
 \\ }}	% Article title
\author{%
\normalsize 
M.\ B.\ Chadwick, M.\ W.\ Paris, and B. M.  Haines
\\[-0.5ex] 
\normalsize Los Alamos National Laboratory, Los Alamos, NM 87545 \\[-0.5ex] 
}
\date{ } % Leave empty to omit a date
\renewcommand{\maketitlehookd}{%
\begin{abstract}
\vspace*{-0.5in}
\noindent {\normalfont\textbf{Abstract}\textemdash}
\noindent 
 In big bang nucleosynthesis (BBN), the deuterium-tritium (DT) fusion reaction, D(T,n)$\alpha$, enhanced by the 3/2$^+$  resonance, is
   responsible for 99\% of primordial $^4$He. This has been known for decades  and has been well documented in the scientific literature.
   However, following  the tradition adopted by authors of learned articles, it was stated in a matter-of-fact manner and not emphasized;
   for most people, it has remained unknown.  This helium became a source for the subsequent creation of $\geq$25\% of the carbon and other
   heavier elements and, thus, a substantial fraction of our human bodies. (To be more precise than $\geq$25\% will require future
   simulation studies on stellar nucleosynthesis.)

Also, without this resonance, controlled fusion energy would be beyond reach. For example, for inertial confinement fusion (ICF),  laser
   energy delivery for the National Ignition Facility (NIF) would have to be approximately 70 times larger for ignition.

Because  the resonance enhances the DT fusion cross section a hundredfold, we propose that the 3/2$^+$ $^5$He excited state be referred to
   as the ``Bretscher state'' in honor of the Manhattan Project scientist who discovered it, in analogy with the well-known 7.6 MeV ``Hoyle
   state'' in $^{12}$C that allows for the resonant 3$\alpha$ formation. 
   
\begin{center}
\par
\noindent Endorsement:  Joyce Guzik, XTD
\vspace*{-0.05in}
\end{center}
\end{abstract}
}

\begin{document}

\maketitle	% Print the title

%----------------------------------------------------------------------------------------
%	Start of text
%----------------------------------------------------------------------------------------

\begin{figure}[htbp]
%\begin{center}
\centering
%\vspace{-1.cm}
\includegraphics[width=2.5in]{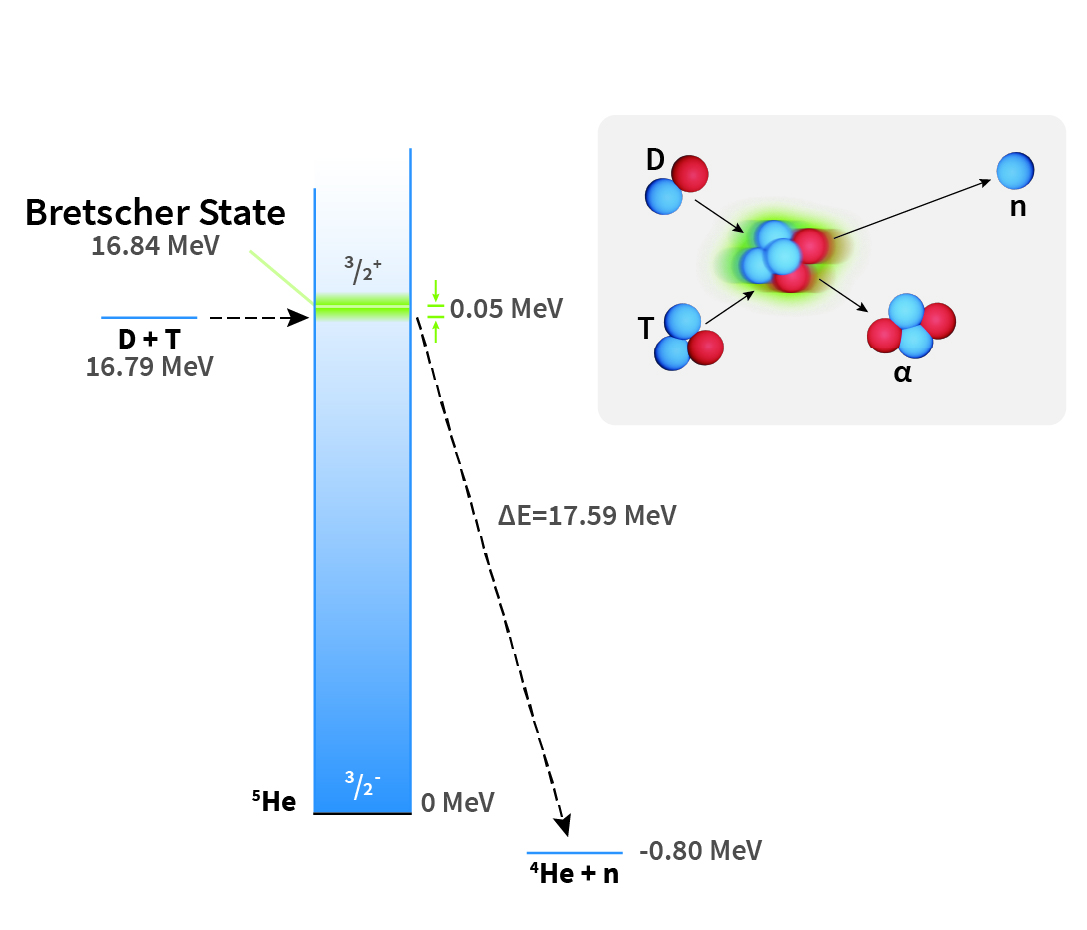}
\vspace{-0.5cm}
\caption{Energy levels in the A=5 system for the D(T,n)$\alpha$ reaction. Resonant enhancement can occur when the resonance (here, 16.84~MeV for the 3/2$^+$) is similar to the D+T separation energy (16.79 MeV). Credit: S.Tasseff.
}
\label{fig:levels}
%\end{center}
\vspace{-0.5cm}
\end{figure}
%\vspace{0.5cm}

\begin{figure}[htbp]
   \centering
   \includegraphics[width=2.6in]{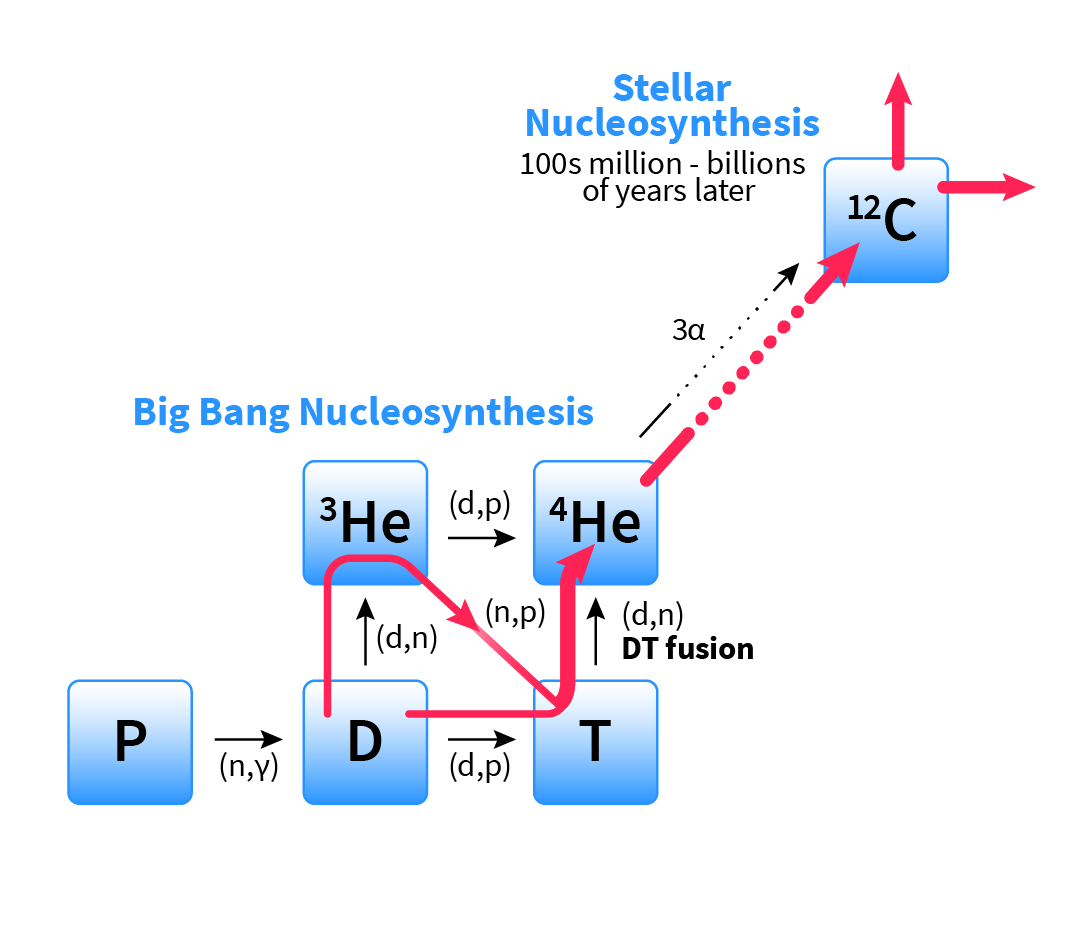}
   \vspace{-0.7cm}
   \caption{A schematic of the dominant BBN pathways and the subsequent  triple-alpha carbon formation in stars. The red curve shows the
   dominant path, which goes through the DT fusion reaction.  Credit: S.Tasseff.}
   \label{fig:pathway}
   \vspace{-0.5cm}
\end{figure}

We recently gave an account\cite{Chadwick:2023} of the earliest DT fusion discoveries, including the important Bretscher \textit{et al.}\
1945--46 Los Alamos measurements \cite{Bretscher:1946a,Bretscher:1949tdn} of the DT cross section down to applications-relevant energies
($\sim$10 keV); an expanded version is under preparation \cite{Chadwick:2022}.  Even though a surprisingly large DT cross section had been
observed earlier in the 1943 Purdue measurement commissioned  by the Manhattan Project, it was Bretscher who was able to make sufficiently
accurate  measurements around the resonance (at 65 keV in the center of mass) so as to identify, characterize, and document
it.\cite{Bretscher:1946a,Bretscher:1949tdn} See Fig.~\ref{fig:levels}. The resonance was found to result in a hundredfold increase in the DT
cross section. In a follow-on paper, Bates and Chadwick\cite{Bates:2022} described the (substantially beneficial) implications of this DT
resonance enhancement for our Los Alamos applications. Here, we extend these discussions to astrophysical nucleosynthesis and ICF
considerations.

%\nopagebreak
 Work in the 1960s and 1970s by Wagoner, Fowler, Hoyle, and Peebles convincingly showed that a standard BBN model accounts for the
 primordial abundances of the light elements, roughly 75\% hydrogen ($^1$H), 25\% helium ($^{4}$He) by mass, and trace amounts of other
 nuclides, created within the first few minutes after the Big Bang by the processes depicted in Fig.~\ref{fig:pathway}. More recently, work by
 Smith \textit{et al.}\cite{Smith:1993} provided a detailed sensitivity study to determine the most important nuclear reaction pathways
 explored in the epoch of the Big Bang, covering the period of time when the universe was about a second old to a few minutes, using
 accurate representations of thermonuclear cross sections, consistent with current understanding. The present account is based on their
 calculations, which are supported by more recent, detailed studies such as those of Grohs \textit{et al.}\cite{Grohs:2016tr}. In order to
 commence our discussion of the production of $^4$He in BBN and the dominance of the DT reaction and the importance of the Bretscher state,
 we first address the precursor, deuterium reactions.

%\newpage
{\bf BBN deuterium production via P(n,$\gamma$)D}. At the early stages of the Big Bang, the temperature was sufficiently high for neutron
and proton number densities to be equal since $n\leftrightarrow p$ transmutations, mediated by the weak interaction, maintained their
equilibrium.  Subsequent cooling, driven by the Hubble expansion of the universe, reduced the $n/p$ ratio (see Figure 2 of
Smith \textit{et al.}\cite{Smith:1993}) as the temperature fell below about 10 billion degrees, equivalent to the mass difference of the neutron and proton
($\sim 1$ MeV).  As the universe further expanded and the temperature dropped below a billion degrees (100 keV) deuterium, created by the
radiative capture reaction P(n,$\gamma$)D, was no longer dissociated in the reverse photodisintegration reaction D($\gamma$,n)P at a rate
sufficient to maintain $n/p$ equilibrium.  The deuterium abundance then grew rapidly to about 1/1000 that of hydrogen. (BBN production of
deuterium differs from its production in stars where there are no free neutrons. Stellar nucleosynthesis of deuterium relies on the
relatively much slower weak interaction $p+p\to$ D$+e^+ +\nu_e$.)

{\bf BBN tritium and $^{3}$He production via DD}. Subsequently, deuterium-deuterium (DD) fusion reactions occurred, as shown in
Fig.~\ref{fig:pathway}.  D(D,p)T and D(D,n)$^3$He fusion reactions occurred at nearly equal rates and created the $A=3$ elements, tritium
(T or $^3$H) and helium-3 ($^3$He). The latter were transmuted by $^3$He(n,p) reactions. These same DD reactions were first observed in
the lab in 1934 by Oliphant, Harteck, and Rutherford, and the cross sections were measured at the Met Lab in Chicago (1942--1944) and at Los
Alamos by Bretscher \cite{Chadwick:2023}.

{\bf BBN $^4$He production via DT}. The calculations of Smith \textit{et al.} (see their Fig.~7a) show that 99\% of the $^{4}$He was
produced via D(T,n)$^4$He, labeled ``DT fusion'' in Fig.~\ref{fig:pathway}, through the large resonant enhancement of this reaction due to
the $^5$He 3/2$^+$ Bretscher state. The remaining 1\% of primordial $^4$He came from the D($^{3}$He,p)$^4$He reaction, which benefits from
the same mirror 3/2$^+$ resonance but is suppressed because of the larger Coulomb repulsion between D and $^3$He.  In fact, much of this
$^3$He was rapidly converted to tritium by the (n,p) reaction, which also underwent DT fusion. $^{4}$He production peaks when the universe's
temperature fell to $\sim$ 70 keV or just below 10$^9$ kelvin, close to the peak of the DT cross section resonance.\footnote{For a 70 keV
early universe BBN temperature,  we find that the peak of the DT fusion reactivity occur at center of mass energy of 65 keV (the peak
resonant energy of the DT cross section), versus 35 keV for  a   10 keV temperature ICF burn. Also, the DT fusion rate at  a 70 keV BBN
temperature is about five times higher than at 10 keV.}  The importance of DT in BBN was noted by Hupin \textit{et al.}\cite{Hupin:2019} in their
interesting paper on \textit{ab initio} cross section calculations.  The central point here is that $^{4}$He, the fuel for further
nucleosynthesis of heavier elements in stars, which constitutes about 25\% by mass of the baryonic matter in the universe, was dominantly
made via resonant DT fusion in BBN.

{\bf Stellar nucleosynthesis}. When the first generation of stars formed, they  were formed from primordial material: 25\% $^4$He and 75\%
$^1$H by mass. The proton-proton (p-p) chain and the carbon-nitrogen-oxygen (CNO) cycle are routes to make more $^4${He} in stars, with the
former being important in lighter stars (below 1.2~M$_\odot$) and the latter in heavier ones (above 1.2~M$_\odot$). This can take the helium
fraction  up to 98--99\% by mass in the stellar core when all the hydrogen is converted. 
The process of igniting the $^4${He} and
starting to produce carbon via the triple-alpha fusion process occurs in stars twice as massive as the sun as they evolve into red giants,
when their cores contract and heat further\cite{Guzik:2023,Burbidge:1957}.

%Joyce Guzik is our local expert on stellar physics. Joyce says: ``In the Sun and other stars, all of the $^1$H in the core is converted to
%$^{4}$He by either the p-p chain (Sun %and stars below 1.2 Msun) or CNO-cycle (stars above 1.2 Msun), so starting from 25\% $^4$He in the
%core, by the end you have 98-99\% $^4$He by mass (with the rest %heavier elements).  Then only after the stars go through the red giant
%branch and their cores contract and heat further do they ignite $^4$He in their cores and start to produce %$^{12}$C via the triple-alpha
%process.''

The triple-alpha reaction is resonantly enhanced by the 7.6~MeV $0^+$ $^{12}$C Hoyle state. Then, more $^4${He} was created  by the CNO
cycle, and heavier element nucleosynthesis continued. Although these stellar nucleosynthesis reactions increased the $^4$He content in the
universe, the increase is small compared with the initial BBN amount.

It is interesting to ask what percentage of the $^4$He that is converted to carbon and heavier elements came from BBN-produced helium compared to stellar-nucleosynthesis-produced helium, where the hydrogen was transmuted to helium through p-p chain and CNO processes. Complete hydrogen burning in stellar nucleosynthesis places a rigorous lower bound of $\ge$25\% $^4$He from BBN, required by the fact that the number of neutrons and protons is conserved. This important point bears reiteration: if the primordial material from which carbon is synthesized had all of its hydrogen converted to helium, then one would have the lower limit, 25\% BBN helium. Otherwise, with incomplete burning of the primordial hydrogen, the percentage of $^4$He would be higher. To be more precise than this will require future studies and simulations.

Therefore, many features of the universe we inhabit were produced through the resonance enhanced DT fusion reaction, notably $\geq$25\% of of
the universe's mass being  $^4$He. Also, at least 25\% of the mass in elements comprising 90\% of our body's mass (the non-hydrogen component) were 
created in BBN DT fusion through the Bretscher state -- a substantial fraction!

If the Bretscher state did not exist, how would our universe be different? This counterfactual universe, which will be addressed in a separate paper, may change the primordial abundances of light elements but would likely not be too dissimilar to our own universe for the heavier elements.
%\cite{Inprepartion}.

{\bf Controlled fusion energy}.  On a related note, we have calculated the impact of the Bretscher state on ICF fusion at the NIF. We did this by reducing the DT cross section by a factor of 100 (since we have shown\cite{Chadwick:2022} that by removing the Bretscher state,  DT is reduced by this amount, becoming similar to the DD fusion reaction).
To achieve the recent megajoule-class energy production results\cite{NIF:2022}, we calculate that one would need a laser drive energy that is roughly 60 times larger than the current 2 MJ delivered. To reach ignition (more energy produced than delivered) would require still more, perhaps 70 times. (See Ref.~\citen{Haines:2017} for a description of how the implosion calculations are done.) This would be an unattainable goal to pursue at present. It is for this same reason that DD plasma burn is not considered feasible anytime soon in controlled fusion energy technologies.

Even a 20\% reduction in the DT fusion cross section has substantial detrimental impacts, and current ICF designs become more marginal, losing about a half of their yield according to our calculations. This 20\% is a historically relevant number because, between the early 1945--46 experiments and more accurate results obtained by 1952, the measured DT cross section values increased such that the calculated reaction rate
at a 10 keV temperature increased by 20\%\cite{Chadwick:2022}.  Recent measurements of NIF implosions are demonstrating that we still may not fully understand DT fusion\cite{Hartouni:2023}.

%\newpage

\begin{figure}[htbp]
\begin{center}
%\vspace{-1.cm}
\includegraphics[width=2.5in]{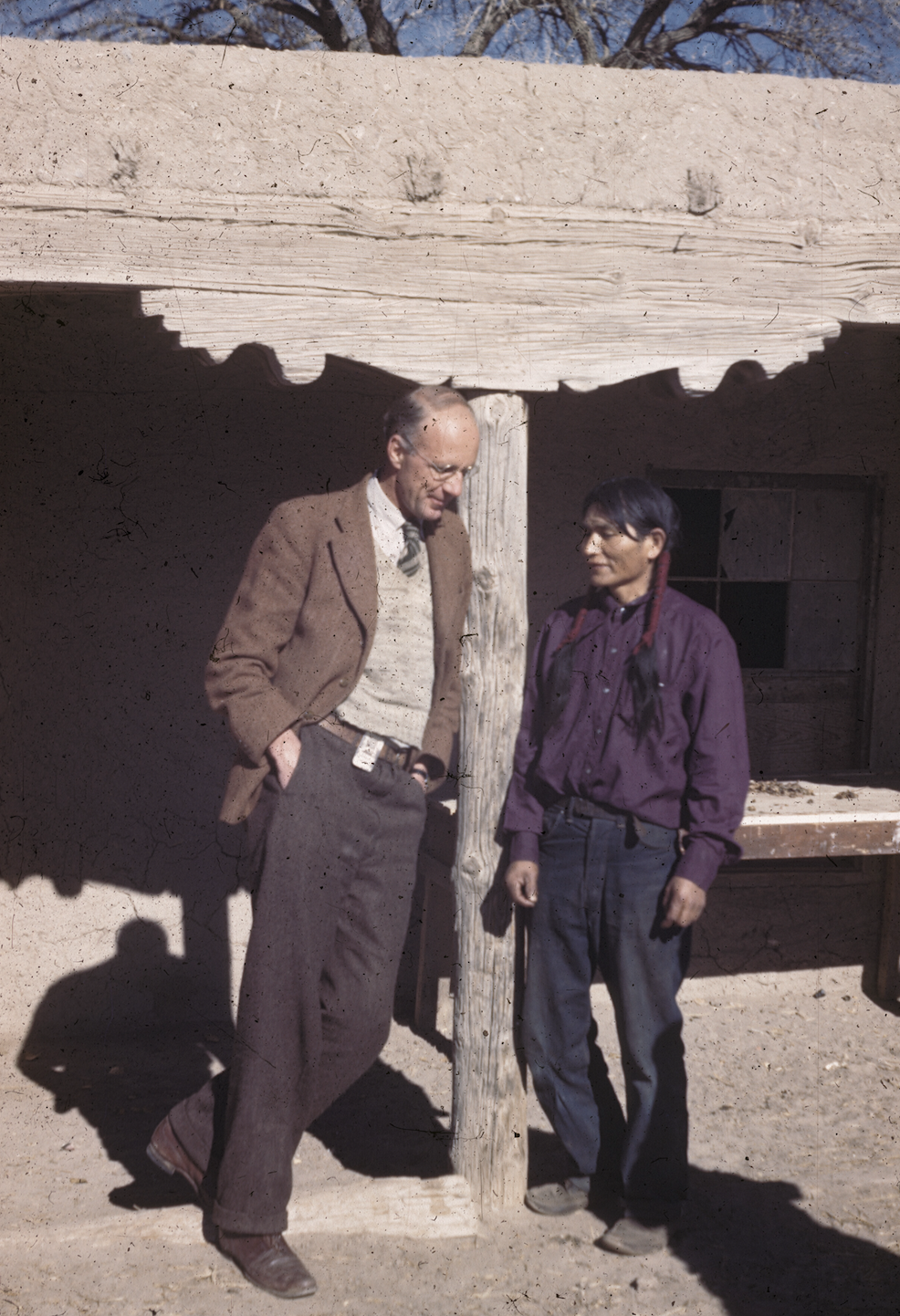}
\caption{Egon Bretscher at the San Ildefonso Pueblo, NM, January 1946 (courtesy Churchill Archives Center, Cambridge University).}
\label{fig:bretscher-pic}
\end{center}
%\vspace{-1.0cm}
\end{figure}

{\bf The Bretscher state?} Because the $^{5}$He 3/2$^+$ 16.8~MeV state discovered by Bretscher\cite{Bretscher:1946a,Bretscher:1949tdn}  is responsible for the fast rate of the DT reaction and the concomitant implications described in this short paper, we propose that the nuclear science community refer to it as the ``Bretscher state,'' in analogy with the $^{12}$C Hoyle state. See Fig.~\ref{fig:bretscher-pic} for a photo of the Swiss-British Egon Bretscher taken around the time that he measured the DT cross section.

\begin{figure}[htbp]
\begin{center}
%\vspace{-1.cm}
\includegraphics[width=2.5in]{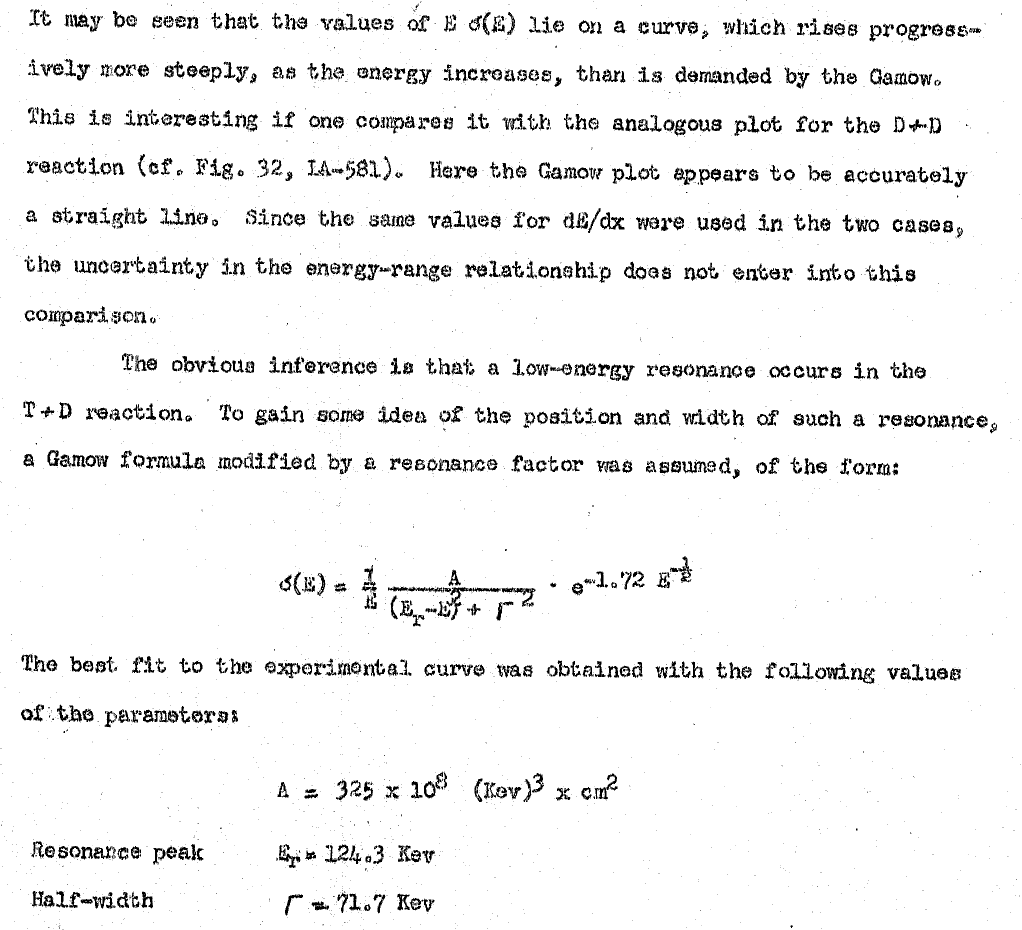}
\caption{An extract from Bretscher and French's paper LA-582\cite{Bretscher:1946a}, February 15, 1946, providing the first identification of the DT resonance.}
\label{fig:bretscher-extract}
\end{center}
%\vspace{-1.0cm}
\end{figure}

Los Alamos' archives hold Fermi's F-Division monthly progress reports, and from these we see that Bretscher and French obtained their first DT data in July 1945 (for a cm energy of 16 keV).
A detailed characterization of the DT resonance was given in Bretscher's Los Alamos report LA-582, dated February 15, 1946\cite{Bretscher:1946a}. An image extract is given in Fig.~\ref{fig:bretscher-extract}, and one can see that this document is identical to the 1949 version published later in the {\it Physical Review}\cite{Bretscher:1949tdn}. This 1946 Los Alamos report was initially classified as a secret document. Its open publication was delayed until 1949. The report provides two approaches to determine the resonance energy. One (see Fig.~\ref{fig:bretscher-extract}) gave a triton incident energy of 124 keV, another gave 343 keV. In the center of mass, these values are equivalent to 50 and 137 keV, respectively, spanning the known value today of 65 keV. Although Bretscher was the first to document and quantify the resonance's energy, we think it very likely that Bethe would have recognized the likely presence of a resonance as early as 1943 when the first DT cross section data were obtained at Purdue\cite{Chadwick:2023,Chadwick:2022} (though at energies above the resonance's peak). The spin of the resonance was correctly identified by 1952---see our discussion on the evolving understanding of the 3/2$^+$ resonance in Ref.~\citen{Chadwick:2022}.

%Pretty much nobody today has heard of Bretscher, so this may be a tough sell. His 1945-46 measurements, which were published in the Physical Review in %1949\cite{Bretscher:1949tdn}, have  been cited only 54 times. But then again, we have pointed out that the first-ever observation of DT fusion, in 1938 by Ruhlig, has never %been cited until we pointed this out\cite{Chadwick:2023}! It is the case that the impact of Bretscher's work, described here, has been underappreciated. 

While Bretscher is generally credited with correctly anticipating the properties of $^{239}$Pu,\cite{Chadwick:2022,Chadwick:2021ANS} his discovery of the DT resonance has gone largely unappreciated. His 1945--46 measurements, which were published in the {\it Physical Review} in 1949, \cite{Bretscher:1949tdn}  have been cited only 54 times, compared with the 460 citations of Hoyle's paper. This is similar---strangely---to the fact that the first-ever observation of DT fusion, in 1938 by Ruhlig,\cite{Ruhlig:1938} had never been cited until our recent paper!\cite{Chadwick:2023} It is with this context that we propose the naming of the 3/2$^+$ DT resonance as the ``Bretscher state,'' in the spirit of its being anthropically important, like the Hoyle state.

 In Ref.~\citen{Chadwick:2022}, we present a brief biography of Bretscher, which includes his early work at Cambridge, his time at Los Alamos, his friendship with Bloch and Staub, and his later leadership roles in nuclear energy research at Harwell.

Again, we note that the importance of DT fusion for BBN is not a new insight\cite{Smith:1993}. This short letter aims to communicate this remarkable fact to those in our Los Alamos community who think about DT fusion for a living.
   
\vspace{0.25in}

{\bf Acknowledgments.} We acknowledge useful discussions with Michael Smith (a BBN expert who enthusiastically endorsed the importance of this underappreciated story and provided many valuable comments), Jonathan Katz and Joyce Guzik (who educated us on stellar nucleosynthesis), Michael Bernardin, Gerry Hale, Craig Carmer and Tom Kunkle. This article is released as Los Alamos report LA-UR-23-23800 (2023).

%------------------------------------------------

%----------------------------------------------------------------------------------------
%	References
% 		ADD your bib filename at the top - default name is test.bib
%----------------------------------------------------------------------------------------
\vspace{0.25in}
\noindent\rule{0.35\textwidth}{.4pt}

%\bibliographystyle{}
%\small\bibliography{bibliography-mbc.bib}  % Add your bib file here are manually add your references

%% LaTeX2e file `test.bib'
%% generated by the `filecontents' environment
%% from source `article' on 2020/10/01.
%

%\bibliographystyle{plain}
%\bibliographystyle{unsrt}
%\bibliographystyle{ans_js}   
%\bibliographystyle{ans}   
%\small\bibliography{master}  % Add your bib file here are manually add your references

\bibliographystyle{ans_js}   
\small\bibliography{bibliography.bib}  % Add your bib file here are manually add your references

\end{document}